# Tactics and Tallies: Inferring Voter Preferences in the 2016 U.S. Presidential Primaries Using Sparse Learning[*]

Yu Wang[†]
Political Science
University of Rochester
Rochester, NY, United States

Yang Feng, Jiebo Luo
Computer Science
University of Rochester
Rochester, NY, United States

Xiyang Zhang
Psychology
Beijing Normal University
Beijing, China

## ABSTRACT
In this paper, we propose a web-centered framework to infer voter preferences for the 2016 U.S. presidential primaries. Using Twitter data collected from Sept. 2015 to March 2016, we first uncover the tweeting tactics of the candidates and then exploit the variations in the number of 'likes' to infer voters' preference. With sparse learning, we are able to reveal neutral topics as well as positive and negative ones.

Methodologically, we are able to achieve a higher predictive power with sparse learning. Substantively, we show that for Hillary Clinton the (only) positive issue area is women's rights. We demonstrate that Hillary Clinton's tactic of linking herself to President Obama resonates well with her supporters but the same is not true for Bernie Sanders. In addition, we show that Donald Trump is a major topic for all the other candidates, and that the women's rights issue is equally emphasized in Sanders' campaign as in Clinton's.

Lessons from the primaries can help inform the general election and beyond. We suggest two ways that politicians can use the feedback mechanism in social media to improve their campaign: (1) use feedback from social media to improve campaign tactics within social media; (2) formulate policies and test the public response from the social media.

## 1. INTRODUCTION
Twitter is playing a significant role in connecting the presidential candidates with voters [17]. Candidates increasingly formulate issue policies and attack rival candidates over Twitter. Some of the candidates' tweets have even entered into the Democratic and the Republican debates.[1] Between September 18, 2015 and March 1st, 2016, Hillary Clinton posted 1973 tweets, Bernie Sanders 2375 tweets, Donald Trump 3175 tweets, Ted Cruz 1876 tweets, and Marco Rubio 1333 tweets.[2] These tweets constitute a valuable data source because they are explicitly political in nature, they are many, and, importantly, they carry feedback information from the voters in the form of 'likes.'

In this paper, we solve two problems. We first study the tweeting tactics of the candidates: we analyze which political figures are mentioned in these tweets and what issues are raised. We then use L1-regularized negative binomial regression to infer voters' preference over these politicians and issues. Our study focuses on the five major candidates during the primaries: Hillary Clinton (D), Bernie Sanders (D), Donald Trump (R), Ted Cruz (R), Marco Rubio (R).[3]

Figure 1 shall illustrate our points well. It presents three tweets that Donald Trump (R) posted on February 9th, 2016, all of which are political in nature. The first tweet talks about drugs. The second tweet raises the issue of ObamaCare and points towards President Obama (D). The third tweet is about the ISIS. Trump supporters responded to these three tweets differently, assigning to the third tweet the most 'likes' and to the second tweet the fewest 'likes.' By connecting topics with responses, we are therefore able

---

---

[1]For example, during the Democratic debate in Flint, Michigan, a tweet by Bernie Sanders targeting Hillary Clinton became the focal point. During the eleventh Republican debate, Donald Trump explicitly invited the audience to check his Twitter account.

[2]We do not count retweets, because retweets do not have as a feature the number of 'likes' and our focus is on the number of 'likes' in this work.

[3]The selection is based on both polling results and on the number of delegates that each candidate has. Marco Rubio (R) dropped out of the race on March 15th, 2016. Throughout, we follow the convention that Republican candidates are marked with (R) and Democratic candidates are marked with (D).

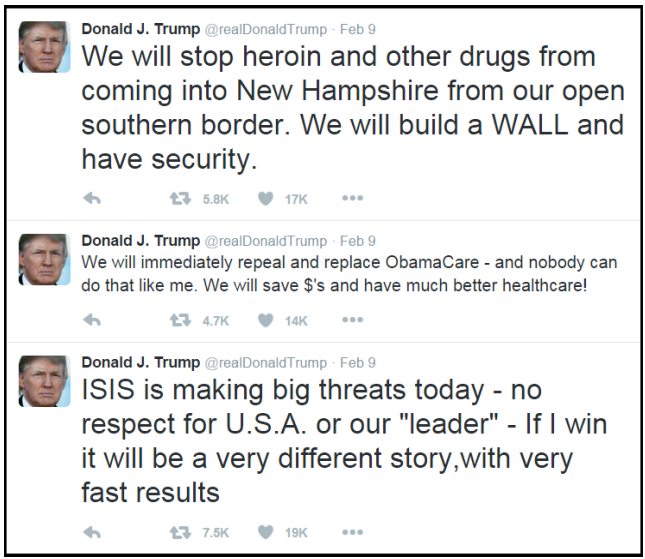

Figure 1: Selected tweets Donald Trump (R) posted on February 9th.

to infer the preferences of the voters.

## 2. RELATED WORK

Our work builds upon previous research in electoral studies, behavioral studies, and sparse learning.

A number of studies have found that campaign and news media messages can alter voters' behavior [13, 8]. According to Gabriel S. Lenz, public debates help inform some of the voters about the parties' or candidates' positions on the important issues [10]. In our work, we assume that tweets posted by the presidential candidates reveal their policy positions in various dimensions and that supporters reveal their policy preference by deciding whether or not to 'like' the tweets.

There are a large body of studies on using social media data to analyze and forecast election results. DiGrazia et al. [3] find a statistically significant relationship between tweets and electoral outcomes. MacWilliams [11] suggests that a candidate's number of 'likes' in Facebook can be used for measuring a campaign's success in engaging the public. According to Williams and Gulati [22], the number of Facebook fans constitutes an indicator of candidate viability. Wang, Li and Luo [20] study the growth pattern of Donald Trump's followers. Gayo-Avello, Metaxas and Mustafaraj [5] advocate that research should be accompanied with a model explaining the predicative power, which we heed to.

Substantively, our paper is closely related to three existing studies of the 2016 U.S. presidential election. Wang, Li, and Luo [19] use Twitter profile images to study and compare the demographics of the followers of Donald Trump and Hillary Clinton. Wang et. al. [18] analyze the demographic differences between the followers and unfollowers. Wang, Li, and Luo [21] model the number of 'likes' that each Trump tweet receives. Our work uses both the number of candidate followers (as a control variable) and the number of 'likes', and we study all the five major candidates rather than focus exclusively on Donald Trump.

There are quite a few studies modeling individual behaviors on social media. Lee et al. [9] model the decision to retweet, using Twitter user features such as agreeableness, number of tweets posted, and daily tweeting patterns. Mahmud, Chen, and Nichols [12] model individuals' waiting time before replying to a tweet based on their previous replying patterns. Wang, Li, and Luo [21] use Latent Dirichlet Allocation (LDA) [1, 2] to extract tweet topics and model follower preferences. In this paper, we search for specific topics using predefined keywords. Compared with LDA, the limitation of our approach is that topics present in the document all share the same weight. The advantage of our approach is that our labels are more definitive and objective.

One of our innovations is to apply L1-regularization to uncover neutral topics. In implementing the L1-regularized negative binomial regression, our work borrows heavily from [14], which introduces stochastic coordinate descent methods for LASSO [16] and for logistic regression. William Greene [6] discusses the likelihood formulation of negative binomial regression. Hastie et al. [7] give a very good discussion on using cross validation to select the penalty term and on using bootstrapping to draw inferences. To the best of our knowledge, our study is the first to implement and apply the regularized negative binomial regression to a real world problem.

## 3. DATA

We use the dataset *US2016*, constructed by us with Twitter data.[4] The dataset contains a tracking record of the number of followers for all the major candidates in the 2016 presidential race, including Hillary Clinton (D), Bernie Sanders (D), Donald Trump (R), Ted Cruz (R), and Marco Rubio (R).[5] The dataset spans the entire period between September 18th, 2015 and March 1st, 2016 and is updated every 10 minutes.

Our dataset *US2016* contains all the tweets that the five candidates posted during the same period and the number of 'likes' that each tweet has received. In Table 1, we report the summary statistics of the dependent variable: 'likes.'

Table 1: Summary statistics

| Variable | Mean | Std. Dev. | Min. | Max. | N |
|---|---|---|---|---|---|
| Donald Trump | 4677.911 | 3742.242 | 722 | 32636 | 3175 |
| Ted Cruz | 548.552 | 608.731 | 8 | 8209 | 1876 |
| Marco Rubio | 590.479 | 879.356 | 3 | 11731 | 1333 |
| Hillary Clinton | 1800.052 | 1636.849 | 119 | 19923 | 1973 |
| Bernie Sanders | 2979.268 | 2966.524 | 204 | 57635 | 2375 |

We refer to these 'likes' as tallies and, in line with [11], we assume that the more likes the better. To visualize these tallies, we plot the density distribution for each candidate, grouped by party affiliation. We align the x axis so that it is easy to compare the distribution both across candidates and across parties. We observe that in the Democratic party, Sanders' tweets tend to receive more 'likes' than Clinton's tweets. Among Republican candidates, Trump's tweets receive more 'likes' than Cruz and Rubio. Equally important,

---
[4]Some of the studies based on this dataset include [21, 20, 19].
[5]Eleven other candidates, including John Kasich (R) and Martin O'Malley (D), are also included in the dataset.

we observe large variations in the distribution for all candidates.

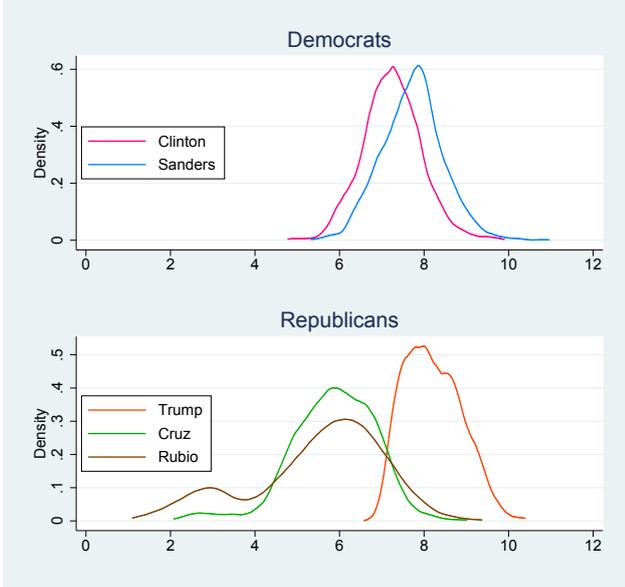

Figure 2: Distribution of Likes in Log Units

We believe part of the variations can be attributed to the topics embedded in the tweets: a more preferred topic generates more 'likes.' To operationalize this idea, we first multi-label each tweet for the following individual-based topics: President Obama (D), Hillary Clinton (D), Bernie Sanders (D), Martin O'Malley (D), Donald Trump (R), Ted Cruz (R), Marco Rubio (R), Jeb Bush (R), Ben Carson (R), Rand Paul (R), John Kasich (R), and Chris Christie (R).[6] We then multi-label each tweet for issue-based topics: ISIS, immigration, Iran, women's rights, education, drugs, gun control, abortion, economy and the Wall Street.[7]

Topic features are binary. We derive these features using keyword matching. For example, we assign to the Obama topic 1 for tweets that contain "Obama" and assign 1 to the Rubio topic for tweets containing "marcorubio" (case-sensitive) or "Rubio." For issue topics, we first transform the tweets to lowercase before the matching procedure. For example, we assign 1 to the abortion topic for tweets that either contain "abortion" or "planned parenthood."

In addition to topic feature variables, we control for the number of followers, the length of the tweet (after removing stop words), and whether or not the tweet contains an http link. In Figure 3, we report time-series follower data for all the five candidates during the primaries. One immediate observation is that Hillary Clinton (D) and Donald Trump (R) dominate their respective party in terms of Twitter followers.

---

[6]By the end of the New Hampshire primary, Martin O'Malley, Rand Paul, and Chris Christie have quit the race.
[7]The selection of political figures is based on the poll performance. For poll data, please refer to http://elections.huffingtonpost.com/pollster#2016-primaries. The selection of political issues follows the Bing Political Index. Available at https://blogs.bing.com/search/2015/12/08/the-bing-2016-election-experience-how-do-the-candidates-measure-up.

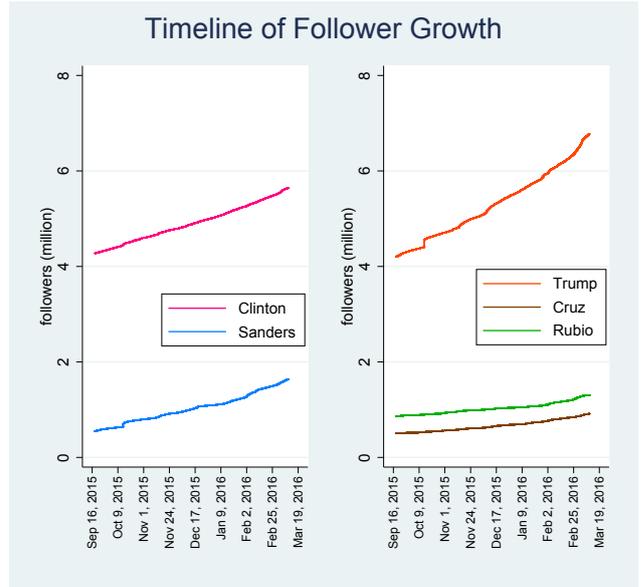

Figure 3: Number of Followers of the Candidates, from September 2015 to March 2016.

## 4. METHODOLOGY

In this section, we first report on how we formulate our problem as an L1 regularized loss minimization problem and then we detail the coordinate descent algorithm for solving the problem, and the parameter selection procedure.

### 4.1 Model Formulation

Our problem starts as a standard negative binomial regression problem, linking the number of 'likes', which is count data, to the explanatory topics. In this regression, the conditional likelihood of the number of 'likes', $y_j$, is formulated as

$$f(y_j|v_j) = \frac{(v_j\mu_j)^{y_j} e^{-v_j\mu_j}}{\Gamma(y_j + 1)}$$

where $\mu_j = exp(\mathbf{x_j}\boldsymbol{\beta})$ is the link function that connects our topics to the number of 'likes' in the tweets and $v_i$ is a hidden variable with a Gamma($\frac{1}{\alpha}$, $\alpha$) distribution.[8] After plugging in the topic variables, the loss function, which is the negative unconditional log-likelihood of the 'likes' takes the form:

$$L = -\sum_{j=1}^{N}[ln(\Gamma(m+y_j)) - ln(\Gamma(y_j+1)) - ln(\Gamma(m))$$
$$+ m\,ln(p_j) + y_j ln(1-p_j)]$$
$$p = 1/(1+\alpha\mu)$$
$$m = 1/\alpha$$
$$\mu = exp(\beta_0 + \beta_1\text{Follower Count} + \beta_2\text{Tweet Length}$$
$$+ \beta_3\text{Hyperlink} + \beta_4\text{Self Referencing}$$
$$+ \boldsymbol{\beta_4} \cdot \textbf{Political Figures} + \boldsymbol{\beta_5} \cdot \textbf{Political Issues})$$

where $\alpha$ is the over-dispersion parameter and will be estimated as well. Now we combine loss $L$ with a penalty for the L1 norm of the coefficients $\boldsymbol{\beta}$ and arrive at the final formulation of our optimization problem.

---

[8]For a detailed introduction to the formulation of the negative binomial likelihood, please see [15, 6].

$$\min_{\boldsymbol{\beta},\alpha} L(\alpha, \langle \boldsymbol{\beta}, \mathbf{X} \rangle, \mathbf{Y}) + \lambda ||\boldsymbol{\beta}||_1$$

## 4.2 Coordinate Descent Algorithm

Solving this minimization problem using coordinate descent, we first calculate the derivative of L with respective to $\boldsymbol{\beta}$ and $\log(\alpha)$ as follows:[9]

$$\frac{\partial L}{\partial \beta_i} = -\sum_{j=1}^{N}[(y_i - m)\frac{1}{1+\alpha\mu_j}\mu_j x_i + \frac{y_j}{\mu_j}\mu_j x_i]$$

$$\frac{\partial L}{\partial \ln(\alpha)} = -\sum_{j=1}^{N}[\frac{1}{\alpha^2}(\psi(m) - \psi(m+y_j) - \ln(p_j)) + [-(m+y_j)\frac{1}{1+\alpha\mu_j}\mu_j + \frac{y_j}{\alpha}]]\alpha$$

where $\psi(x)$ is the digamma function, $\psi(x) := \frac{\partial \ln(\Gamma(x))}{\partial x}$.

We apply regularization to all the independent variables but not to the over-dispersion parameter $\alpha$. Our coordinate descent algorithm is detailed as follows:

---
Algorithm: Coordinate Descent for L1 Regularized Negative Binomial[10]

---
## p: number of features, $\eta$: learning rate;
let $\boldsymbol{\beta}=\mathbf{1} \in R^p$, $\alpha = 1 \in R$, $\eta = 0.002$
**for** iter=1,2,...N do
  **for** i=1,2, ..,p do
    $L'_i = \frac{\partial L}{\partial \beta_i}$
    if $\beta_i - \eta L' > \eta\lambda$:
      $\beta_i \leftarrow \beta_i - \eta L' - \eta\lambda$
    else if $\beta_i - \eta L' < -\eta\lambda$:
      $\beta_i \leftarrow \beta_i - \eta L' + \eta\lambda$
    else:
      $\beta_i \leftarrow 0$
  **end for**
  $L'_\alpha = \frac{\partial L}{\ln(\alpha)}$
  $\ln(\alpha) \leftarrow \ln(\alpha)$-$\eta\ln'_\alpha$
  $\alpha \leftarrow e^{ln(\alpha)}$
**end for**

---

## 4.3 Model Selection

Following suggestions in [7], we use cross validation to select the penalty term and the model. Specifically, we use 5-fold cross validation, i.e. iteratively 20% of the dataset is held out for validation, to select $\lambda_{CV}$ that minimizes the averaged mean squared prediction error (MSE):

$$MSE = \frac{1}{N}\sum_{j=1}^{j=N}(y_j - e^{\mathbf{x}_j \hat{\mathbf{b}}})^2$$

---
[9]We calculate the derivative of L with respect to $\log(\alpha)$ instead of $\alpha$ to ensure that $\alpha$ stays positive throughout the optimization procedure.
[10]We have posted our codes at http://sites.google.com/site/wangyurochester.

where $y_j$ is the true number of 'likes' for the $j$th tweet and $e^{\mathbf{x}_j \hat{\mathbf{b}}}$ is the model's prediction.

When $\lambda_{CV}$ equals zero, i.e. no penalty, our model is equivalent to the standard negative binomial regression, so we treat the $\lambda_{CV} = 0$ as the baseline model.[11] We report the cross validation error curve in Figure IV. Based on cross validation and the MSE metric, the model we use for later point estimation and inference yields the highest predictive power.

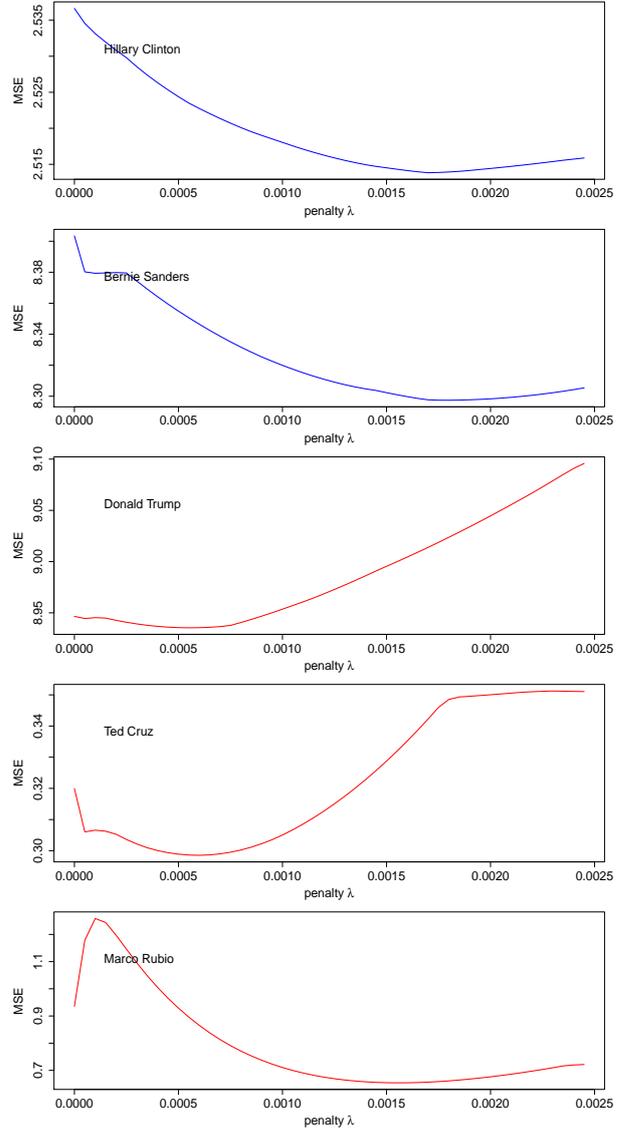

**Figure 4: Penalty Selection for Each Presidential Candidate**

In drawing inferences on $\beta$'s distribution, we use 1,000 bootstrap realizations of $\hat{\beta}_{\lambda_{cv}}$ [4]. The detailed results are reported in Section 5.

---
[11]When the penalty term is zero, our results are identical to the standard outputs from Stata (http://www.stata.com) and R.

## 5. EMPIRICAL RESULTS

In this section, we present our main findings. We first report on the candidates' tweeting tactics and then on our estimation of voters' preference. Our presentation and discussion will focus mainly on Hillary Clinton and Donald Trump as they have won the presidential nomination of their respective party and will compete against each other in the general election.

### 5.1 Tactics

In Figure 5, we report on the frequency with which each of the political figures has been mentioned by the Democratic candidates.[12] We find Clinton focuses mostly on Donald Trump (R) and President Obama (D) and that Bernie Sanders focuses on Hillary Clinton and Donald Trump. Their varied emphasis on President Obama is consistent with the view that Hillary Clinton has the implicit backing from the President.

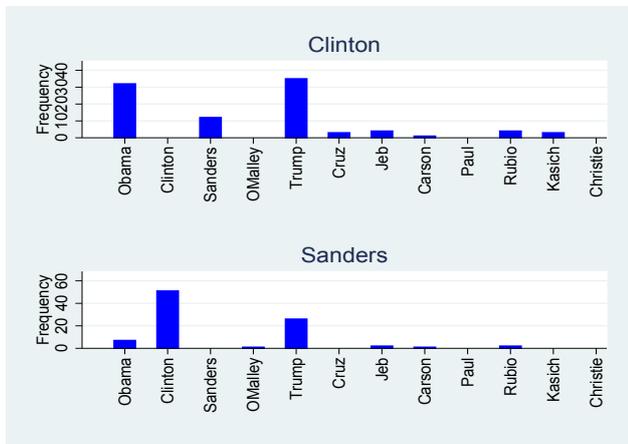

Figure 5: Political Figures Mentioned by Hillary Clinton and Bernie Sanders.

In Figure 6, we report on the frequency that each of the political figures has been mentioned by the Republican candidates. We find Trump focuses mostly on Ted Cruz (R), Jeb Bush (R) and Hillary Clinton (D). Cruz focuses on Donald Trump, President Obama and Hillary Clinton. Marco Rubio focuses almost exclusively on the Democrats: President Obama and Hillary Clinton.

In Figure 7, we report on the distribution of policy issues for the Democratic candidates. We observe that Hillary Clinton focuses heavily on women's rights issue and on gun control, while Sanders talks about the economy and attacks the Wall Street the most.

In Figure 8, we report on the distribution of policy issues for the Republican candidates. Compared with the Democrats, the Republicans focus more on foreign policies: ISIS, immigration and Iran and less on domestic issues. For Donald Trump, the most prominent issue is immigration.

### 5.2 Voter Preference

In this section, we present our estimation of voters' preferences. We obtain the penalty terms for each candidate using

---

[12] We do not count self-referencing in our calculation, i.e., Clinton is not considered a political topic for Clinton herself.

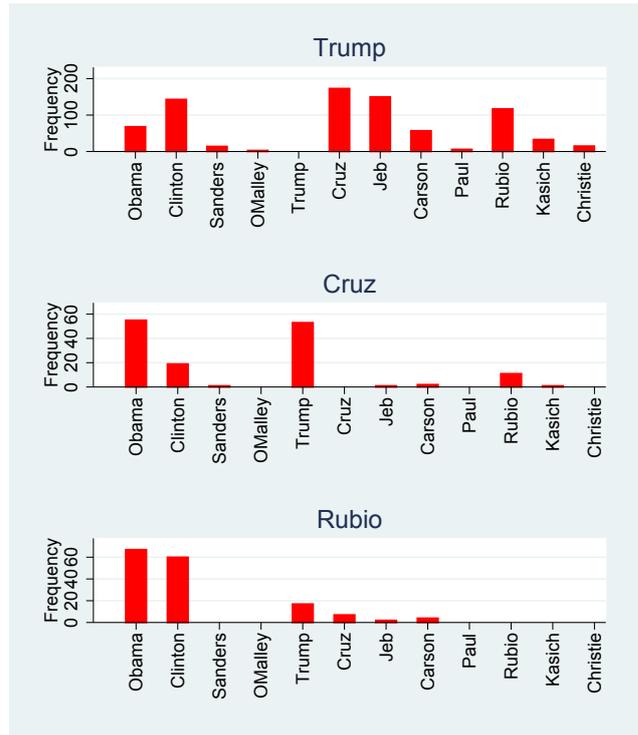

Figure 6: Political Figures Mentioned by Trump, Cruz and Rubio.

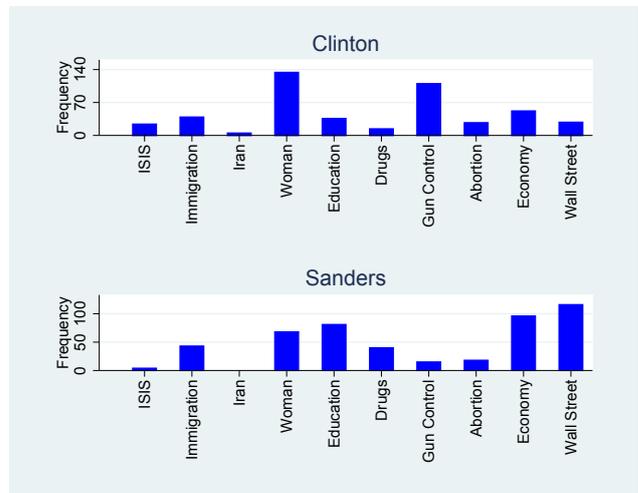

Figure 7: Policy Issues Raised by Hillary Clinton and Bernie Sanders.

5-fold cross validation. We use bootstrapping to calculate the standard errors for each topic coefficient. The distributions of the coefficients for each candidate are reported at the end of the paper (Figure 9 - Figure 13).

We report the estimated coefficients for Donald Trump in the first column of Table 2. We find that Trump receives more 'likes' when he attacks Democrats and fewer 'likes' if he attacks fellow Republicans. This is consistent with the findings reported in [21], which uses unsupervised LDA. Trump is also particularly strong on ISIS. Due to regularization, Rand Paul (R) and abortion prove to be neutral topics.

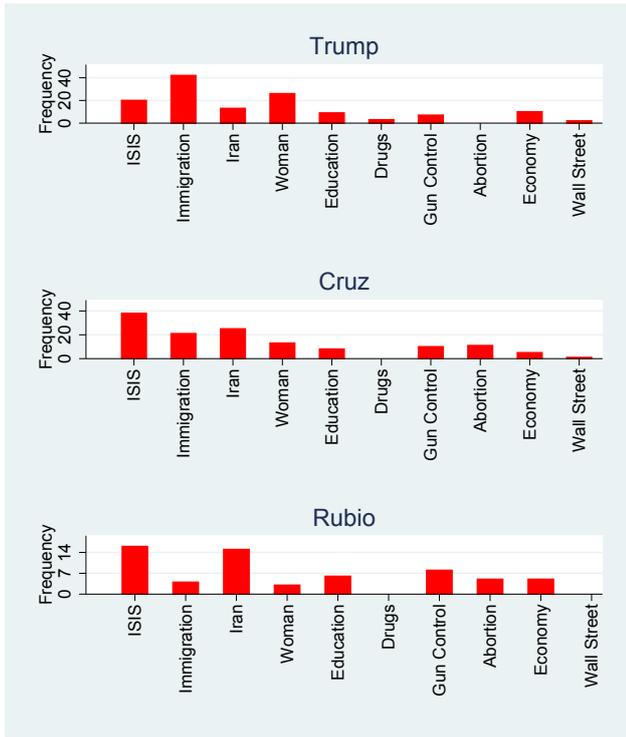

**Figure 8: Policy Issues Raised by Trump, Cruz and Rubio.**

By contrast, Hillary Clinton, reported in the last column of Table 2, clearly benefits from her association with President Obama. Clinton receives more 'likes' on women's rights and fewer 'likes' on the economy and on Wall Street. The penalty term for Clinton (0.0017) is larger than that for Trump (0.00055). As a result, more coefficients for Hillary Clinton turn out to be zeros.

For Ted Cruz (R), he receives more 'likes' when attacking Hillary Clinton (D) and Donald Trump (R). The coefficients for other political figures are either zero or not significant. In issue areas, Cruz is strong on ISIS but not others. For Marco Rubio (R), the only two topics that earn him more 'likes' are Donald Trump and ISIS. His attacks on President Obama and Hillary Clinton do not resonate with his supporters very well. Bernie Sanders (D) is known as the candidate calling for a political revolution, and the preferences of his voters do show peculiarity.

We observe that Sanders is the only candidate who receives more 'likes' in the area of education and fewer 'likes' on immigration.[13] This is consistent with the candidate's policies: Sanders is the first candidate to propose free college education (which Clinton has copied later), and Sanders, whose father immigrated from Poland, supports a path to U.S. citizenship for the undocumented immigrants.

Focusing on the parties, we observe that the Republicans

---

[13] For a review of Sanders' rise and his proposed policies, please see http://www.nytimes.com/2016/03/13/opinion/sunday/the-bernie-sanders-revolution.html.

**Table 2: L1 Regularized Negative Binomial Regression**

|  | Trump | Cruz | Rubio | Sanders | Clinton |
|---|---|---|---|---|---|
| *Likes* | | | | | |
| Constant | -2.068** | -1.791** | -2.111** | 0.810** | **0.000** |
|  | (0.105) | (0.088) | (0.13) | (0.106) | (0.167) |
| Followers | 6.532** | 6.653** | 5.793** | 3.828** | 1.787** |
|  | (0.176) | (0.315) | (0.598) | (0.731) | (0.333) |
| Length | 0.131** | 0.454** | 0.514** | **0.000** | -0.052 |
|  | (0.022) | (0.045) | (0.086) | (0.039) | (0.049) |
| Http | 0.097** | 0.143 | 0.134 | -0.417** | -0.171** |
|  | (0.024) | (0.079) | (0.118) | (0.045) | (0.045) |
| Obama | 0.386** | 0.073 | -0.127 | **0.000** | 0.445** |
|  | (0.09) | (0.086) | (0.078) | (0.045) | (0.186) |
| Clinton | 0.236** | 0.698** | 0.057 | 0.225 | -0.378** |
|  | (0.046) | (0.332) | (0.079) | (0.126) | (0.037) |
| Sanders | 0.122 | **0.000** |  | 0.319 | 0.128 |
|  | (0.126) | (0.095) |  | (0.215) | (0.14) |
| Omalley | 0.253 |  |  | **0.000** |  |
|  | (0.164) |  |  | (0.0) |  |
| Trump | -0.311** | 0.789** | 1.704** | 0.961** | 0.336 |
|  | (0.022) | (0.14) | (0.19) | (0.145) | (0.196) |
| Cruz | -0.098** | -0.307** | 0.210 |  | 0.000 |
|  | (0.037) | (0.047) | (0.413) |  | (0.009) |
| Jeb | -0.085** | **0.000** | **0.000** | **0.000** | **0.000** |
|  | (0.036) | (0.212) | (0.142) | (0.0) | (0.009) |
| Carson | -0.073 | 0.169 | **0.000** | 0.185 | **0.000** |
|  | (0.058) | (0.248) | (0.241) | (0.326) | (0.007) |
| Rand | **0.000** |  |  |  |  |
|  | (0.087) |  |  |  |  |
| Rubio | -0.072 | 0.345 | -0.560** | **0.000** | **0.000** |
|  | (0.042) | (0.277) | (0.221) | (0.218) | (0.055) |
| Kasich | -0.048 | **0.000** |  |  | **0.000** |
|  | (0.068) | (0.141) |  |  | (0.123) |
| Christie | -0.007 |  |  |  |  |
|  | (0.184) |  |  |  |  |
| ISIS | 0.414** | 0.306** | 0.725** | **0.000** | **0.000** |
|  | (0.098) | (0.095) | (0.231) | (0.025) | (0.051) |
| Immigration | 0.111 | -0.053 | **0.000** | -0.257** | -0.045 |
|  | (0.092) | (0.123) | (0.108) | (0.071) | (0.086) |
| Iran | 0.089 | -0.105 | **0.000** |  | **0.000** |
|  | (0.106) | (0.123) | (0.04) | (—) | (0.118) |
| Women's rights | -0.008 | 0.060 | **0.000** | 0.389** | 0.176** |
|  | (0.094) | (0.172) | (0.051) | (0.083) | (0.058) |
| Education | 0.382 | -0.090 | **0.000** | 0.221** | **0.000** |
|  | (0.33) | (0.157) | (0.036) | (0.062) | (0.083) |
| Drugs | 0.012 |  |  | **0.000** | **0.000** |
|  | (0.261) |  |  | (0.066) | (0.081) |
| Gun Control | 0.571 | **0.000** | **0.000** | 0.468 | 0.018 |
|  | (0.292) | (0.075) | (0.044) | (0.297) | (0.071) |
| Abortion |  | **0.000** | 0.276 | 0.235 | 0.047 |
|  |  | (0.103) | (0.224) | (0.149) | (0.07) |
| Economy | **0.000** | **0.000** | **0.000** | -0.253** | -0.229** |
|  | (0.056) | (0.097) | (0.043) | (0.048) | (0.071) |
| Wall Street | -0.348 | **0.000** |  | -0.080 | -0.330** |
|  | (0.185) | (0.065) |  | (0.058) | (0.074) |
| $\alpha$ | 0.0700 | 0.0016 | 0.0274 | 0.0708 | 0.1930 |
| $\lambda$ | 0.00055 | 0.00065 | 0.00155 | 0.0018 | 0.0017 |

Standard errors in parentheses
** $p < 0.05$
Zero coefficients in bold.

benefit from their position on ISIS, while ISIS is a neutral topic for the Democrats. The women's rights issue wins Democratic candidates more 'likes,' but it is neutral for the Republicans. The economy is apparently hurting the Democratic candidates but not the three Republican candidates. The results also show that voters do not like to see candidates attack fellow party members.

For the control variables, we observe that the more followers one has the more 'likes' one receives, which is intuitive. In terms of the length of the tweets, we find that longer tweets tend to help Republicans win more 'likes,' but not so for the Democrats. Lastly, tweets with a hyperlink tend to receive more 'likes' for Donald Trump. The opposite is true for the Democrats.[14]

Lastly, we emphasize that our regularized negative binomial regression outperforms the standard negative binomial, i.e. with no penalty terms, in two dimensions. First, we are able to achieve a smaller prediction error, which can be seen in Figure 4. Second, interpretation of the results is made substantially easier as many (33, to be specific) of the coefficients have shrunk to zero. As a result, campaigns are better able to focus on the winning issues.

## 6. CONCLUSIONS

Twitter is playing an important role in connecting presidential candidates and potential voters through the tweeting and 'likes' channels. In this paper, we have proposed a framework to infer voter preferences via this feedback mechanism. Using Twitter data collected from September 2015 to March 2016, we first uncovered the tweeting tactics of the candidates and then we exploited the variations in the number of 'likes' to infer voters' preference. Besides positive and negative topics, we were also able to reveal neutral topics with sparse learning.

Methodologically, we were able to achieve a higher predictive power with sparse learning. Substantively, we showed that for Hillary Clinton the (only) positive issue area is women's rights. We demonstrated that Hillary Clinton's tactic of linking herself to President Obama resonates well with her supporters but the same is not true for Bernie Sanders. In addition, we showed that Donald Trump is a major topic for all the other candidates and that the women's rights issue is equally emphasized in Sanders' campaign as in Clinton's.

Lessons from the primaries can help inform the general election and beyond. We suggest two ways that politicians could use the feedback mechanism in social media and more generally on the web to improve their campaign: (1) use feedback from social media to improve campaign tactics within social media; (2) formulate policies and test the public response from the social media.

## 7. ACKNOWLEDGMENTS

Yu Wang would like to thank the Department of Political Science at the University of Rochester for generous funding. We also gratefully acknowledge support from the University, New York State through the Goergen Institute for Data Science, and our corporate sponsors Xerox and Yahoo.

---

[14] Images constitute an integral part of the candidates' campaign strategy. As future research, we are examining how the posted images, their contents more specifically, affects followers' reaction.

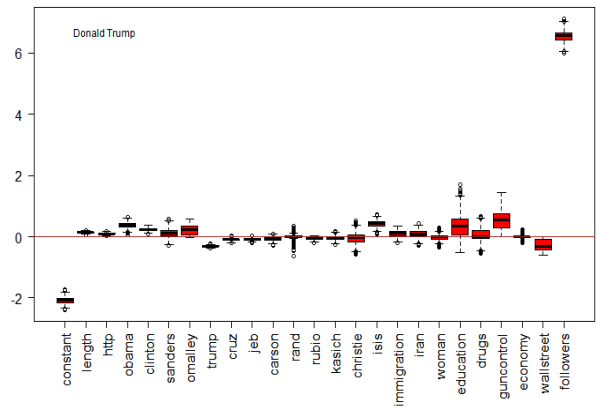

Figure 11: Coefficient Estimation for Donald Trump.

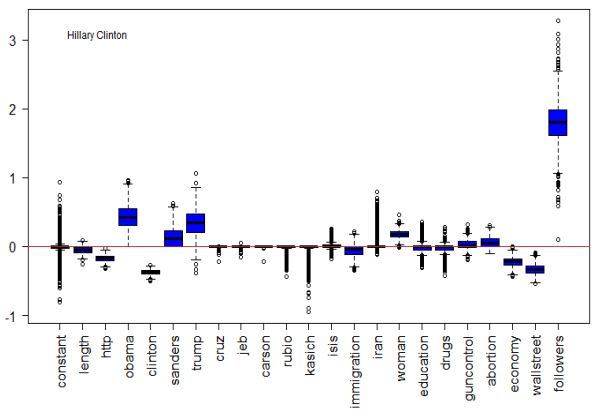

Figure 9: Coefficient Estimation for Hillary Clinton.

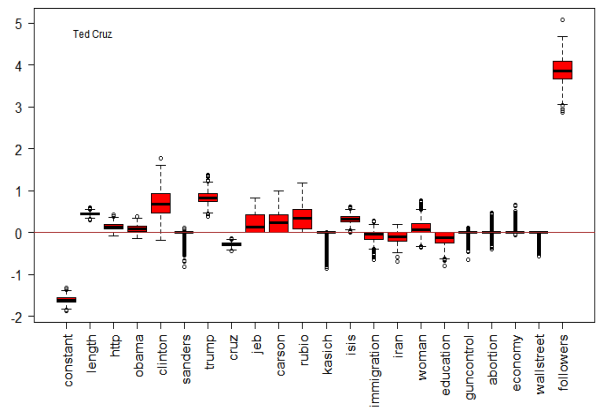

Figure 12: Coefficient Estimation for Ted Cruz.

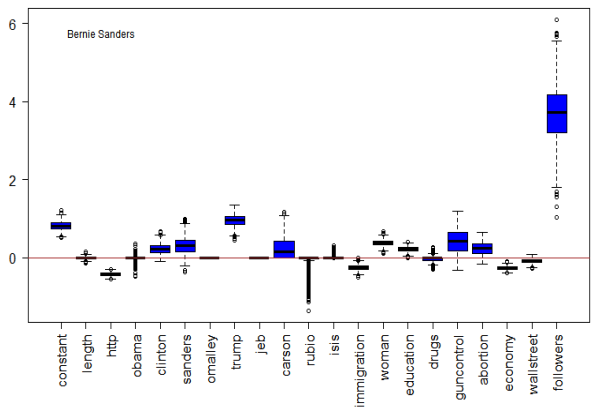

Figure 10: Coefficient Estimation for Bernie Sanders.

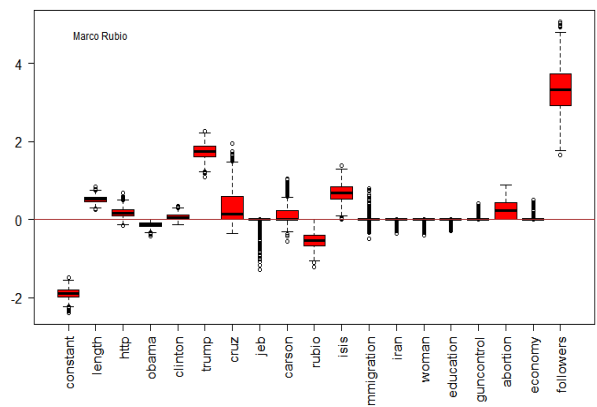

Figure 13: Coefficient Estimation for Marco Rubio.